\documentclass[10pt]{iopart}	
\usepackage{amssymb,bm}
\usepackage{calrsfs}
\usepackage {graphicx}

\usepackage{hyperref}
\hypersetup{
    colorlinks=true,        
    linkcolor=blue,          
    citecolor=blue,        
    filecolor=magenta,      
    urlcolor=blue           
}
\usepackage{color} 				
\definecolor{dgray}{rgb}{0.6,0.6,0.6}
\definecolor{dmag}{rgb}{0.6,0.0,0.6}
\usepackage[normalem]{ulem} 	

\newcommand{\eqref}[1]{{(\ref{#1})}}

\begin{document}

\title[Re-examining the quadratic approximation ...]{Re-examining the quadratic approximation in theory of a weakly interacting Bose gas with condensate: the role of nonlocal interaction potentials}

\author{M S Bulakhov$^{1,2}$, A S Peletminskii$^{1}$, S V Peletminskii$^{1}$, Yu~V~Slyusarenko$^{1,2}$ and A G Sotnikov$^{1,3}$}
\address{$^{1}$ Akhiezer Institute for Theoretical Physics, NSC KIPT, 61108 Kharkiv, Ukraine}
\ead{aspelet@kipt.kharkov.ua, a\_sotnikov@kipt.kharkov.ua}
\address{$^{2}$ Karazin Kharkiv National University, 61022 Kharkiv, Ukraine}
\address{$^{3}$ Institute of Solid State Physics, TU Wien, 1020 Vienna, Austria}

\submitto{\jpb}

\date{\today}

\begin{abstract}
We derive and analyze the coupled equations of quadratic approximation of the Bogoliubov model for a weakly interacting Bose gas. The first equation determines the condensate density as a variational parameter and ensures the minimum of the grand thermodynamic potential. The second one provides a relation between the total number of particles and chemical potential. Their consistent theoretical analysis is performed for a number of model interaction potentials including contact (local) and nonlocal interactions, where the latter provide nontrivial dependencies in momentum space. We demonstrate that the derived equations have no solutions for the local potential, although they formally reproduce the well-known results of the Bogoliubov approach. At the same time, it is shown that these equations have the solutions for physically relevant nonlocal potentials. We show that in the regimes close to experimental realizations with ultracold atoms, the contribution of the terms originating from the quadratic part of the truncated Hamiltonian to the chemical potential can be of the same order of magnitude as from its $c$-number part. Due to this fact, in particular, the spectrum of single-particle excitations in the quadratic approximation acquires a gap. The issue of the gap is also discussed.
\end{abstract}


\maketitle

\section{Introduction}

While the phenomenon of Bose-Einstein condensation (BEC) in an ideal gas was predicted in 1925 \cite{Einstein}, the path to its experimental evidence in dilute vapors of alkali atoms took over 70 years \cite{Anderson,Davis, Bradley}. These first experiments triggered a large number of new perspective and unique theoretical and experimental investigations \cite{Dalfovo,Pethick,Stringari}. Now, ultracold atomic gases provide remarkable opportunities to study and model various effects and phenomena in quantum many-body systems in a well controlled manner.

Theoretical description of BEC is usually based on the original Bogoliubov microscopic theory \cite{Bogoliubov}  or non-linear time-dependent Gross-Pitaevskii equation \cite{Gross,Pitaevskii} with a trapping potential of experimental interest. Both of approaches are applicable for dilute weakly-interacting Bose gases and describe, respectively, their homogeneous and inhomogeneous states at zero temperature. While the Gross-Pitaevskii approach has proved to be a successful tool in describing the inhomogeneous structures relevant to experiments (e.g., vortices, solitons, density profiles, and breathing modes \cite{Dalfovo,Pethick,Stringari}), the specific Bogoliubov  microscopic theory has a more rigorous mathematical formulation and is well justified from the statistical physics point of view. The latter has also been generalized to study equilibrium spatially inhomogeneous states (periodic structures) \cite{PPSl} and to derive the kinetic equations for the quasiparticle distribution function and the condensate density \cite{PSS,KD}, as well as hydrodynamic equations of superfluids \cite{PSSc,LPS,KirkDorf}.

The main ingredient of the Bogoliubov theory is a replacement of creation and annihilation operators of the condensate particles by $c$-numbers in all operators of relevant physical quantities. This procedure is supposed to be exact in the infinite-volume limit and intuitively clear, since a macroscopic number of particles is condensed into the single state and, therefore, one can neglect the non-commutativity of the corresponding creation and annihilation operators. However, a more rigorous justification of this procedure is not so trivial. The first attempt of such a justification was also given by Bogoliubov, who introduced the concept of quasi-averages \cite{Bog} and related the $c$-number replacement to spontaneous breaking of the U(1) symmetry. Later, it was proved that under sufficiently general conditions for the two-body interaction potential, the Bogoliubov replacement gives the exact result for pressure in the infinite-volume limit \cite{Ginibre}. Recently, the role of $c$-number substitution and the gauge symmetry breaking in theory of BEC has been extensively discussed in a number of studies \cite{Lieb,Suto1,Suto2,ZB,LSSY}. In parallel, along with the bosonic version of the Hartree-Fock-Bogoliubov and $\Phi$-derivable approximations reviewed in Ref.~\cite{Andersen2004}, other schemes not involving the $c$-number formalism were developed to describe the BEC phase (see, e.g., Refs.~\cite{Kras1993, PhysRep1994,Bobrov2010,CMP2013,Bobrov2016,Ett,Poluektov2017}).

By applying the $c$-number replacement of creation and annihilation operators of particles with zero momentum in the pair-interaction Hamiltonian, Bogoliubov proposed to take into account only the $c$-number terms and those that are quadratic in creation and annihilation operators of particles with nonzero momentum \cite{Bogoliubov}. In the case of weak interaction, the terms containing three and four operators are inessential and can be neglected. These must be properly taken into account when describing the effects originating from interaction of quasiparticles. Such a truncated Hamiltonian underlies the quadratic approximation that allows one to introduce, in a natural way, the concept of quasiparticles and to compute the basic thermodynamic quantities and the spectrum of single-particle excitations. However, since the U(1) symmetry of the truncated Hamiltonian is broken, it no longer commutes with the particle  number operator. 
Therefore, one can approach the problem by considering the grand canonical ensemble, where the chemical potential, being a Lagrange multiplier, ensures the conservation of the total number of particles. For this system the chemical potential was firstly obtained by Beliaev~\cite{Beliaev} and, subsequently, by Hugenholtz and Pines \cite{Hugenholtz} by using a partial summation of the perturbation series expansion of the Green's functions. They found the non-power-law series expansions for the ground-state energy and the chemical potential in terms of the gas parameter (the interaction was parameterized by the constant scattering amplitude). According to these results, the leading term in the expansion of the chemical potential originates from the $c$-number part of the truncated Hamiltonian and leads to a gapless spectrum of the single-particle excitations. Other terms, including those coming from the quadratic part of the Hamiltonian, are typically much smaller and can be neglected in the leading approximation.
It was also shown that the quasiparticle spectrum can not have an energy gap at zero momentum~\cite{Hugenholtz}. 
However, as it was later pointed out by Pines \cite{Pines}, the correctness of this statement depends on the validity of the expansions into series of the perturbation theory. Moreover, there were also some doubts concerning proper treatment of the depletion effect and disconnected diagrams in the Hugenholtz-Pines perturbative technique \cite{Misawa}.

In physics of cold atomic gases, the Fourier transform of the real interatomic interaction is usually replaced by the constant value characterized by the $s$-wave scattering length \cite{Dalfovo,Pethick,Stringari}. When attempting to compute the chemical potential or the ground-state energy for this interaction by taking into account the corresponding terms from the quadratic part of the truncated Hamiltonian, divergences of the corresponding integrals at large momenta are encountered, so that it is necessary to use the renormalization of the coupling constant \cite{Dalfovo,Pethick,Stringari}. The divergences also appear when calculating the ground-state energy by taking into account the term with four operators in the truncated Hamiltonian \cite{PPS}. Therefore, the replacement of real interaction potentials by their constant values (or by the scattering lengths only) is not so ``inoffensive'' approximation, see also experimental observations \cite{Hazlett2012PRL} and theoretical studies~\cite{Caballero2013} for Fermi systems.

In this paper, we study the general coupled equations of quadratic approximation for a weakly interacting Bose gas with BEC in the case of model momentum-dependent (nonlocal) interaction potentials. The first equation represents the necessary condition ensuring the minimum of the grand thermodynamic potential. The second one determines a relation between the total particle number and the chemical potential. 
The attempts to solve analytically the resulting system of nonlinear equations face with considerable difficulties even within a perturbative approach. In this case, as pointed out in Ref.~\cite{Tolmachev}, the results contain non-analyticity with respect to weak interaction that indicates the absence of an adequate perturbative approach. Therefore, we analyze these equations numerically at zero temperature for two types of non-local interaction potentials: the semi-transparent sphere and Gaussian potentials. The main advantage of these potentials is that they allow to avoid the mentioned divergences in the corresponding integrals due to vanishing at large momenta and, at the same time, they contain a contact (local) interaction as a limiting case. As we show, the obtained results drastically change the generally accepted picture. In particular, it is shown that, in contrast to the conventional treatment (as described above), the contribution of the terms coming from the quadratic part of the Hamiltonian into the chemical potential can be of the same order of magnitude as from the $c$-number part. Therefore, these terms can not be neglected and should be properly taken into account. This fact immediately leads to a gapful spectrum of the single-particle excitations.

Experimentally, there are still considerable limitations in the direct access of the low-momentum region of the quasiparticle dispersion, see, e.g., Refs.~\cite{Steinhauer2002, Beauvois2018}. For this reason, the treatment of experimental data allows for both gapful and gapless structure. However, historically, the most of existing extrapolations involve the assumptions of the phonon-like nature of the spectrum or employ the local character of interatomic interaction, thus {\sl a priori} exclude the possible existence of a gap in the single-particle excitation spectrum. From this point of view, the experimental proof of the gap presence in the single-particle excitation spectrum could indicate that the replacement of complex interatomic potentials by the contact interaction is not sufficient to correctly describe all essential physics of the weakly interacting atomic gases with BEC.

The paper is organized as follows. In Sec.~\ref{sec2} we provide a derivation of the general equations of quadratic approximation for a weakly interacting Bose gas. In Sec.~\ref{sec3} we discuss the model nonlocal potentials and transform the obtained equations to the dimensionless form, convenient for a numerical analysis. Sec.~\ref{sec4} deals with numerical solutions of the derived equations and the corresponding analysis of physical observables. Finally, we summarize our results in Sec.~\ref{sec5}.

\section{General equations of quadratic approximation}\label{sec2}

When studying a normal state of a many-body system of weakly interacting particles, the standard perturbative approach in interaction \cite{Akh-Pel} (or the thermodynamic perturbation theory) is usually employed to find the corrections to the statistical operator, the thermodynamic potential, and the many-particle distribution functions. However, as a rule, the standard technique of perturbative theory becomes inapplicable for the systems with spontaneously broken symmetry, thus a description requires the developments of new asymptotic methods. In particular, the standard perturbative approach fails to describe the equilibrium state of a non-ideal Bose gas with BEC even if the interparticle interaction is weak. This is due to the fact that the divergent terms appear in series of the usual perturbative approach. For this reason, one needs to construct an appropriate theory to describe the condensed phase of a Bose gas that breaks the gauge symmetry.

As mentioned in the introduction, such a specific theory was developed by Bogoliubov \cite{Bogoliubov}. Since the number of condensed particles acquires a macroscopic value, which is proportional to the volume of the system $\mathcal{V}$, it was proposed to treat the creation and annihilation operators of particles with zero momentum as $c$-numbers, so that $a^{\dagger}_{0}$ and $a_{0}$ are replaced by $N_{0}^{1/2}$ in all operators of relevant physical quantities, where $N_{0}\propto{\mathcal V}$ is the number of particles condensed into the state with ${\bf p}=0$. The next step is to truncate the initial pair-interaction Hamiltonian, so that it contains only the $c$-number terms and quadratic terms in creation and annihilation operators. The truncated Hamiltonian, or the so-called Bogoliubov quadratic approximation, allows one, in a consistent way, to introduce the quasiparticles and to compute the basic thermodynamic characteristics of the system.

Now, we shortly remind the Bogoliubov microscopic approach and derive the general equations of quadratic approximation. To this end, consider a many-body system of interacting bosonic atoms described by the following Hamiltonian:
\begin{eqnarray}
H=H_{0}+V=\sum_{\bf p}\varepsilon_{\bf p}\,a^{\dagger}_{\bf p}a_{\bf p}
&+&{1\over 2\mathcal{V}}\sum_{{\bf p}_{1}...{\bf p }_{4}}
\nu({\bf p}_{1}-{\bf p}_{3})\,
\nonumber
\\
&&\times
a_{{\bf p}_{1} }^{\dagger}a_{{\bf p}_{2}}^{\dagger}a_{{\bf p}_{3}}a_{{\bf p}_{4}}\,\delta_{{\bf p}_{1}+{\bf p}_{2},{\bf p}_{3}+{\bf p}_{4}}, \label{eq:2.1}
\end{eqnarray}
where $\varepsilon_{\bf p}=p^{2}/2m$ is the kinetic energy of an atom with the mass $m$, $a_{\bf p}^{\dagger}$ and $a_{\bf p}$ are the creation and annihilation operators that meet the usual bosonic commutation relations, $[a_{\bf p},a_{{\bf p}'}^{\dagger}]=\delta_{{\bf p},{\bf p}'}$ and $[a_{\bf p},a_{\bf p}']=0$, and $\nu({\bf p})$ is the Fourier transform of the potential interaction energy for two atoms. Here, for simplicity, we do not take into account the spin degrees of freedom and exchange interaction between atoms. The generalization of the Bogoliubov approach to include the spin degrees of freedom was studied in Refs.~\cite{APS,PelPelSl}.
The particle number operator
\begin{equation} \label{eq:2.2}
	N=\sum_{\bf p}a^{\dagger}_{\bf p}a_{\bf p}
\end{equation}
commutes with the above Hamiltonian that guarantees the conservation of the total particle number.

Performing the replacement of $a_{0}$ and $a^{\dagger}_{0}$ with $N_{0}^{1/2}$ in Eq. (\ref{eq:2.1}), one obtains
\begin{eqnarray} \label{eq:2.4}
H(N_{0})=H_{0}+f(N_{0})+{\partial f(N_{0})\over\partial N_{0}}N'+N_{0}V^{(2)}
+N_{0}^{1/2}V^{(3)}+V^{(4)},
\end{eqnarray}
where
\begin{equation} \label{eq:2.5}
f(N_{0})={\nu(0)N_{0}^{2}\over 2\mathcal{V}}, \quad N'=\sum_{{\bf p}\neq 0}a^{\dagger}_{\bf p}a_{\bf p},
\end{equation}
and
\begin{eqnarray}
V^{(2)}&=&{1\over \mathcal{V}}\sum_{{\bf p}\neq 0}\nu({\bf p})a^{\dagger}_{{\bf p}}a_{{\bf p}}
+{1\over 2\mathcal{V}}\sum_{{\bf p}\neq 0}\nu({\bf p})\left[a^{\dagger}_{{\bf p}}a^{\dagger}_{-{\bf p}}+a_{{\bf p}}a_{-{\bf p}}\right], 
\nonumber\\
V^{(3)}&=&{1\over \mathcal{V}}\sum_{{\bf p}_{1}...{\bf p}_{3}\neq 0}
\nu({\bf p}_{2})\left[a^{\dagger}_{{\bf p}_{1}}a_{{\bf p}_{2}}a_{{\bf p}_{3}}\delta_{{\bf p}_{1},{\bf p}_{2}+{\bf p}_{3}}+a^{\dagger}_{{\bf p}_{1}}a^{\dagger}_{{\bf p}_{2}}a_{{\bf p}_{3}}\delta_{{\bf p}_{1}+{\bf p}_{2},{\bf p}_{3}}\right], 
\nonumber\\
V^{(4)}&=&{1\over 2\mathcal{V}}\sum_{{\bf p}_{1}...{\bf p}_{4}\neq 0}\nu({\bf p}_{1}-{\bf p }_{4})\,a^{\dagger}_{{\bf p}_{1}}a^{\dagger}_{{\bf p}_{2}}a_{{\bf p}_{3}}a_{{\bf p}_{4}}\,
\delta_{{\bf p}_{1}+{\bf p}_{2},{\bf p}_{3}+{\bf p}_{4}}. \label{eq:2.6}
\end{eqnarray}
Under the $c$-number replacement, the particle number operator is reduced to $N(N_{0})=N_{0}+N'$. Therefore, the Gibbs statistical operator corresponding to the grand canonical ensemble reads
\begin{eqnarray} \label{eq:2.7}
w(N_{0})=\exp[\Omega-\beta\mathcal{H}(N_{0})], 
\quad
\mathcal{H}(N_{0})=H(N_{0})-\mu N_{0}-\mu N',
\end{eqnarray}
where $\beta=1/T$ is the reciprocal temperature, $\mu$ is the chemical potential (or the Lagrange multiplier) that guarantees the conservation of the total particle number. The grand thermodynamic potential $\Omega$ as a function of $\beta$, $\mu$, and $N_{0}$ is determined from the normalization condition $\Tr\left[w(N_{0})\right]=1$, where the trace is taken in the space of occupation numbers of bosons with ${\bf p}\neq 0$. In turn, the number of condensed atoms $N_{0}$ is obtained from the condition for the minimum of the thermodynamic potential,
\begin{equation} \label{eq:2.8}
	{\partial\Omega\over\partial N_{0}}
    =-\beta\bigg\{\mu-\Tr
    \left[w(N_{0}){\partial H(N_{0})\over\partial N_{0}}\right]\bigg\}=0,
\end{equation}
whence it follows that
\begin{equation} \label{eq:2.9}
	\mu=\Tr\left[w(N_{0}){\partial H(N_{0})\over\partial N_{0}}\right].
\end{equation}

It is worth stressing that all the derived relations, including Eq.~(\ref{eq:2.9}), are exact. When obtaining them, we only employed the $c$-number replacement of the corresponding creation and annihilation operators and did not use any perturbative approach. Equation~(\ref{eq:2.9}) should be considered as that for determining $n_{0}$ and not as the definition of the chemical potential $\mu$.

The next step is to truncate the Hamiltonian given by Eq. (\ref{eq:2.4}) in a way to preserve only the $c$-number terms and those that are quadratic in creation and annihilation operators, but to neglect the terms of the third and higher orders in $a_{\bf p}$ and $a_{\bf p}^{\dagger}$. Such a truncated Hamiltonian underlies the so-called quadratic approximation of the model and, after the corresponding diagonalization, allows one to describe the system of weakly interacting particles in the language of free quasiparticles. The higher order terms $N_{0}^{1/2}V^{(3)}$ and $V^{(4)}$ in Eq.~(\ref{eq:2.4}) become relevant when describing the interaction effects between quasiparticles themselves.
Therefore, within the quadratic approximation we have
\begin{eqnarray}
	\mathcal{H}^{(2)}(N_{0})&=&f(N_{0})-\mu N_{0}
\nonumber\\
	&&+\sum_{{\bf p}\neq 0}
	\left[\alpha_{\bf p}a^{\dagger}_{\bf p}a_{\bf p}
	+{1\over 2}\beta_{\bf p}\left(a_{\bf p}^{\dagger}a_{-{\bf p}}^{\dagger}
    +a_{\bf p}a_{-{\bf p}}\right)\right],\label{eq:2.10}
\end{eqnarray}
and, consequently,
\begin{equation}
w(N_{0})\simeq w^{(2)}(N_{0})=\exp[\Omega^{(2)}-\beta\mathcal{H}^{(2)}(N_{0})],
\end{equation}
where
\begin{equation} \label{eq:2.11}
	\alpha_{\bf p}=\varepsilon_{\bf p}-\mu+\nu(0)n_{0}+\beta_{\bf p}, \quad \beta_{\bf p}=\nu({\bf p})n_{0}
\end{equation}
with $n_{0}=N_{0}/\mathcal{V}$ being the condensate density (here and below the superscript ``(2)'' denotes the physical quantities up to the second order in the operators $a_{\bf p}$ and $a^\dag_{\bf p}$). The grand thermodynamic potential $\Omega^{(2)}$ is found from the normalization condition $\Tr\left[w^{(2)}(N_{0})\right]=1$. Note that $\Omega$ in Eq.~(\ref{eq:2.7}) coincides with $\Omega^{(2)}$ in the currently-used approximation. 

The operator $\mathcal{H}^{(2)}(N_{0})$ [and thereby $w^{(2)}(N_{0})$] can be reduced to the diagonal form by the unitary transformation $U$,
\begin{equation} \label{eq:2.12}
U\mathcal{H}^{(2)}(N_{0})U^{\dagger}=\sum_{{\bf p}\neq 0}\omega_{\bf p}a^{\dagger}_{\bf p}a_{\bf p}+\mathcal{E}_{0}^{(2)}, \quad UU^{\dagger}=1,
\end{equation}
where $\omega_{\bf p}$ is the quasiparticle energy and $\mathcal{E}_{0}^{(2)}$ is the ``ground-state energy'' of $\mathcal{H}^{(2)}(N_{0})$ [note that $\mathcal{H}^{(2)}(N_{0})$ is not a true Hamiltonian]. For the diagonalization of $\mathcal{H}^{(2)}(N_{0})$, it is sufficient to restrict ourselves to unitary operators $U$, which mix up the operators $a^{\dagger}_{-{\bf p}}$ and $a_{\bf p}$:
\begin{equation}\label{eq:2.13}
Ua^{\dagger}_{\bf p}U^{\dagger}=\mathcal{U}_{\bf p}a^{\dagger}_{\bf p}+\mathcal{V}_{\bf p}a_{-{\bf p}}, ~~Ua_{\bf p}U^{\dagger}=\mathcal{U}_{\bf p}a_{\bf p}+\mathcal{V}_{\bf p}a^{\dagger}_{-{\bf p}}.
\end{equation}

The introduced new operators must satisfy the same bosonic commutation relations, i.e., the transformation must be canonical. This requirement results in the following relations for $\mathcal{U}_{\bf p}$ and $\mathcal{V}_{\bf p}$:
\begin{equation}\label{eq:2.14}
\mathcal{U}_{\bf p}^{2}-\mathcal{V}_{\bf p}^{2}=1, \quad \mathcal{U}_{\bf p}\mathcal{V}_{-{\bf p}}-\mathcal{V}_{\bf p}\mathcal{U}_{-{\bf p}}=0.
\end{equation}
Applying the standard diagonalization procedure \cite{Bogoliubov}, one finds the quasiparticle spectrum,
\begin{equation}\label{eq:2.15}
\omega_{\bf p}=(\alpha^{2}_{\bf p}-\beta^{2}_{\bf p})^{1/2},
\end{equation}
as well as the functions $\mathcal{U}_{\bf p}$ and $\mathcal{V}_{\bf p}$, which define the canonical transformation $U$,
\begin{equation}\label{eq:2.16}
\mathcal{U}_{\bf p}^{2}={(\alpha_{\bf p}+\omega_{\bf p})^{2}\over{(\alpha_{\bf p}+\omega_{\bf p})^{2}-\beta_{\bf p}^{2}}}, \quad \mathcal{V}_{{\bf p}}^{2}={\beta^{2}_{\bf p}\over{(\alpha_{\bf p}+\omega_{\bf p})^{2}-\beta_{\bf p}^{2}}}.
\end{equation}

In general case, the single-particle excitation spectrum determined by Eqs.~(\ref{eq:2.15}) and (\ref{eq:2.11}) exhibits a gap at ${\bf p}=0$,
\begin{equation}
    \omega_{0}^2=(\nu(0)n_{0}-\mu)(3\nu(0)n_{0}-\mu).
\end{equation}
However, this gap vanishes if the chemical potential satisfies the known Hugenholtz--Pines relation \cite{Hugenholtz}, $\mu=\nu(0)n_{0}$ [$\mu\neq 3\nu(0)n_{0}$, see Sec.~\ref{sec3}]. Note that the issue of the gapful spectrum of single-particle excitations in a many-body Bose system with BEC has a long history (see, e.g., Refs.~\cite{Girardeau1959, Wentzel1960, Luban1962, Hohenberg1965, Kondratenko1975,Yukalov2006}) and it has been revived in recent studies \cite{Suto2008,Bobrov2010,Ett,Poluektov2014}. Below we show that such a gap can exist for non-local interaction potentials.

In order to determine $\mathcal{E}_{0}^{(2)}$, let us introduce a vector $|0\rangle$ corresponding to a pure BEC state, i.e., free of the non-condensate particles, $a_{\bf p}|0\rangle=0$. Then, from Eq. (\ref{eq:2.10}), we have $\langle 0|\mathcal{H}^{(2)}(N_{0})|0\rangle=f(N_{0})-\mu N_{0}$. On the other hand, noting that [see Eq.~(\ref{eq:2.12})]
$$
\mathcal{H}^{(2)}(N_{0})=\sum_{{\bf p}\neq 0}\omega_{\bf p}U^{\dagger}a_{\bf p}^{\dagger}UU^{\dagger}a_{\bf p}U+\mathcal{E}_{0}^{(2)}
$$
and using Eqs.~(\ref{eq:2.13}), one obtains
$$
\langle 0|\mathcal{H}^{(2)}(N_{0})|0\rangle=\sum_{{\bf p}\neq 0}\omega_{\bf p}\mathcal{V}_{\bf p}^{2}+\mathcal{E}_{0}^{(2)}.
$$
Finally, performing the comparison of two results and employing the explicit form of $\mathcal{V}_{\bf p}^{2}$, see Eq.~\eqref{eq:2.16}, we arrive at
\begin{equation}\label{eq:2.17}
\mathcal{E}_{0}^{(2)}=
f(N_{0})-\mu N_{0}+{1\over 2}\sum_{{\bf p}\neq 0}(\omega_{\bf p}-\alpha_{\bf p}).
\end{equation}
The transformation given by Eqs.~(\ref{eq:2.13}) reduces the Gibbs statistical operator corresponding to the quadratic approximation to the diagonal form,
\begin{equation}\label{eq:2.18}
Uw^{(2)}(N_{0})U^{\dagger}=\exp\left[\tilde{\Omega}^{(2)}-\beta\sum_{{\bf p}\neq 0}\omega_{\bf p}a^{\dagger}_{{\bf p}}a_{{\bf
p}}\right],
\end{equation}
where
\begin{equation} \label{eq:2.19}
\tilde{\Omega}^{(2)}=\Omega^{(2)}-\beta\mathcal{E}_{0}^{(2)}=\sum_{{\bf p}\neq 0}\ln\left(1-e^{-\beta\omega_{\bf p }}\right).
\end{equation}
The quasiparticle distribution function is then found to be
\begin{equation} \label{eq:2.20}
f_{\bf p}={\partial\tilde{\Omega}^{(2)}\over\partial{(\beta\omega_{\bf p}})}={1\over{e^{\beta\omega_{\bf p}}-1}}.
\end{equation}

For the subsequent analysis, it is convenient to introduce the potential density ${\cal W}=\Omega/\beta\mathcal{V}$ that is a relativistic invariant and up to a sign coincides with the pressure~$P$ \cite{Pel-Pel,Pel}. Then, from Eq.~(\ref{eq:2.19}) we have
\begin{equation} \label{eq:2.21}
{\cal W}^{(2)}=-P={1\over\mathcal{V}}\left[\mathcal{E}_{0}^{(2)}+{1\over\beta}\sum_{{\bf p}\neq 0}\ln\left(1-e^{-\beta\omega_{\bf p }}\right)\right].
\end{equation}

We are now in a position to derive the general coupled equations describing the equilibrium properties of a weakly interacting gas with BEC in the quadratic approximation. First, let us determine the total particle density, $n=N/\mathcal{V}$. According to Eq.~(\ref{eq:2.2}), it reads
\begin{eqnarray*}
	n&=&n_{0}+{1\over\mathcal{V}}\sum_{{\bf p}\neq 0}{\rm Tr}\,\left[w^{(2)}
    (N_{0})a^{\dagger}_{\bf p}a_{\bf p}\right]
    \\
    &=&n_{0}+{1\over\mathcal{V}}
    \sum_{{\bf p}\neq 0}{\rm Tr}\,\left[Uw^{(2)}(N_{0})U^{\dagger}Ua^{\dagger}_{\bf p}U^{\dagger}Ua_{\bf p}U^{\dagger}\right].
\end{eqnarray*}
Then, using Eqs.~(\ref{eq:2.13})-(\ref{eq:2.14}), (\ref{eq:2.16}) we arrive at
\begin{equation}\label{eq:2.22}
n=n_{0}+{1\over2\mathcal{V}}\sum_{{\bf p}\neq 0}\left[{\alpha_{\bf p}\over\omega_{\bf p}}(2f_{\bf p}+1)-1\right].
\end{equation}

Next, we address Eq.~(\ref{eq:2.9}) ensuring the minimum of the grand thermodynamic potential. In the quadratic approximation, we have
\begin{eqnarray*}
	\mu\simeq{\rm Tr}\,\left[w^{(2)}(N_{0}){\partial H^{(2)}(N_{0})
    \over\partial N_{0}}\right]=
    {\rm Tr}\,\left[Uw^{(2)}(N_{0})U^{\dagger}U{\partial H^{(2)}(N_{0})
    \over\partial N_{0}}U^{\dagger}\right],
\end{eqnarray*}
where $H^{(2)}(N_{0})$ is given by Eq.~(\ref{eq:2.4}) with the higher-order terms $N_{0}^{(1/2)}V^{(3)}$ and $V^{(4)}$ being neglected. Taking into account that
\begin{eqnarray*}
	N_{0}{\partial H^{(2)}(N_{0})\over\partial N_{0}}
    &=&f(N_{0})+\mu N_{0}+\mathcal{H}^{(2)}(N_{0})
    \\
    &&-\sum_{{\bf p}\neq 0}\left[\alpha_{\bf p}-\beta_{\bf p}-\nu(0)n_{0}\right]a^{\dagger}_{\bf p}a_{\bf p},
\end{eqnarray*}
and repeating the steps similar to those that resulted in Eq.~(\ref{eq:2.22}), we obtain (see also Ref. \cite{Tolmachev})
\begin{eqnarray}
	\mu\simeq \nu(0)n_{0}&+&{1\over 2\mathcal{V}}\sum_{{\bf p}\neq 0}\left[\nu(0)+\nu({\bf p})\right]
    \left[{\alpha_{\bf p}\over\omega_{\bf p}}(2f_{\bf p}+1)-1\right]
    \nonumber\\
    &-&{1\over 2\mathcal{V}}\sum_{\bf p}{\nu^{2}({\bf p})n_{0}\over\omega_{\bf p}}(2f_{\bf p}+1).
    \label{eq:2.23}
\end{eqnarray}
The general system of coupled Eqs. (\ref{eq:2.22}) and (\ref{eq:2.23}), as well as Eqs.~(\ref{eq:2.11}) and  (\ref{eq:2.15}), allows to express the chemical potential in terms of the temperature and total particle density, i.e., to determine $n_{0}=n_{0}(n,\beta)$ in the quadratic approximation. Note that the first term in Eq.~(\ref{eq:2.23}) is related to the $c$-number part of $\mathcal{H}^{(2)}(N_{0})$, while the other terms originate from its quadratic part.

At zero temperature, the quasiparticle distribution function $f_{\bf p}$ turns to zero and, consequently, Eqs.~(\ref{eq:2.22}) and (\ref{eq:2.23}) become
\begin{equation}\label{eq:2.24}
	n=n_{0}+{1\over2\mathcal{V}}\sum_{{\bf p}\neq 0}
    \left[{\alpha_{\bf p}\over\omega_{\bf p}}-1\right],
\end{equation}
\begin{eqnarray}
	\mu\simeq \nu(0)n_{0}+{1\over 2\mathcal{V}}\sum_{{\bf p}\neq 0}\left[\nu(0)+\nu({\bf p})\right]
    \left[{\alpha_{\bf p}\over\omega_{\bf p}}-1\right]
    -{1\over 2\mathcal{V}}\sum_{\bf p}{\nu^{2}({\bf p})n_{0}\over\omega_{\bf p}}.
    \label{eq:2.25}
\end{eqnarray}
According to Eqs.~(\ref{eq:2.17}) and (\ref{eq:2.21}), the ground-state thermodynamic potential density is given by
\begin{equation}\label{eq:2.26}
{\cal W}^{(2)}={\nu(0)n_{0}^{2}\over 2}-\mu n_{0}+{1\over 2\mathcal{V}}\sum_{{\bf p}\neq 0}(\omega_{\bf p}-\alpha_{\bf p}).
\end{equation}

In the standard Bogoliubov approximation, valid at zero temperature, the chemical potential is defined by the $c$-number part of the truncated Hamiltonian, $\mu\simeq \nu(0)n_{0}$. This chemical potential satisfies the Hugenholtz-Pines relation \cite{Hugenholtz} and immediately leads to a gapless spectrum of single-particle excitations given by Eq.~(\ref{eq:2.15}). Below, we analyze Eqs.~(\ref{eq:2.24}) and (\ref{eq:2.25}) for the contact interaction potential and for some model potentials with nontrivial dependence of their Fourier transforms on momentum. As we shall see, in the first case, Eq.~(\ref{eq:2.25}) has no solution, although it formally leads to the well-known corrections to the chemical potential and the total particle density in terms of the gas parameter. In the second case, the contribution of the terms originating from the quadratic part of $\mathcal{H}^{(2)}(N_{0})$ [the last two terms in Eq.~(\ref{eq:2.25})] can be of the same order of magnitude as $\nu(0)n_{0}$ and should be taken into account, so that the single-particle excitation spectrum acquires a gap.

\section{Zero-temperature quadratic approximation for model potentials}\label{sec3}

\begin{figure}
\includegraphics[width=0.75\linewidth]{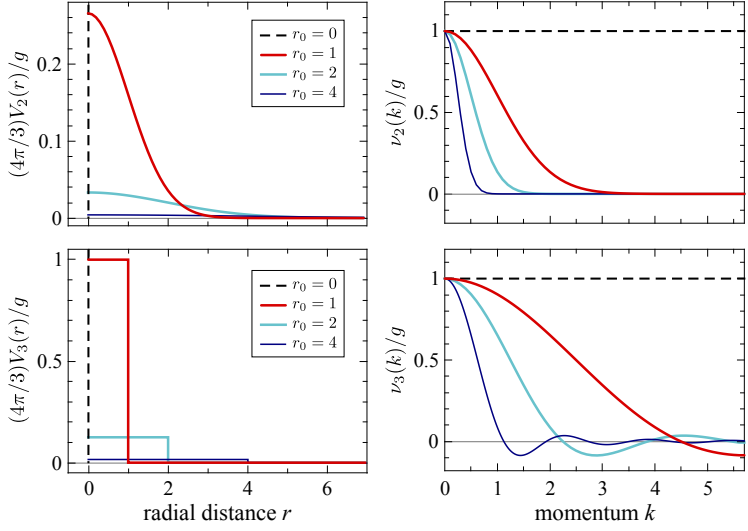}
\caption{Real-space (left) and momentum (right) distributions of pseudopotentials of Gaussian type~$V_2$ (upper row) and semi-transparent spheres~$V_3$ (lower row) at different real-space extents~$r_0$. The local interaction potential $V_1$ and its Fourier transform $\nu_1$ correspond to the $r_0=0$ limits.}\label{fig1}
\end{figure}

To proceed with the analysis of Eqs.~(\ref{eq:2.24})-(\ref{eq:2.26}), it is necessary to specify the Fourier transform of the interaction potential $\nu({\bf p})$,
\begin{equation}\label{eq:nup}
 \nu({\bf p}) = \int  d{\bf r}\, e^{-i{\bf pr}/\hbar} V({\bf r}),
\end{equation}
where $V({\bf r})$ is the two-body interaction potential. Below, we consider three cases of model potentials corresponding to the assumption of spherical symmetry of interactions, see also Fig.~\ref{fig1}.

As the first case, we analyze Eqs.~(\ref{eq:2.24})-(\ref{eq:2.26}) for the contact (local) interaction potential widely used in the physics of ultracold atomic gases to make concrete predictions associated with interaction effects \cite{Dalfovo,Pethick,Stringari}. This potential can be written as $V_1({\bf r})=g\delta({\bf r})$. The corresponding Fourier transform reads
\begin{equation}\label{eq:nup1}
\nu_1({\bf p})=g.
\end{equation}
In this particular case, Eq.~(\ref{eq:2.25}) takes the form
\begin{equation}\label{eq:mu}
	\mu_1=gn_{0}+{g\over \mathcal{V}}
    \sum_{{\bf p}\neq 0}\left[{\alpha_{\bf p}\over\omega_{\bf p}}-{gn_{0}\over 2\omega_{\bf p}}-1\right],
\end{equation}
where, according to Eqs.~(\ref{eq:2.11}) and (\ref{eq:2.15}),
\begin{eqnarray}
	&&\alpha_{\bf p}={p^{2}/2m-\mu+2gn_{0}},
    \nonumber\\
    &&\omega_{\bf p}=\left[\left(p^{2}/2m-\mu+2gn_{0}\right)^{2}
    -g^{2}n_{0}^2\right]^{1/2}.
    \label{eq:spectr}
\end{eqnarray}

\begin{figure}
\includegraphics[width=0.75\linewidth]{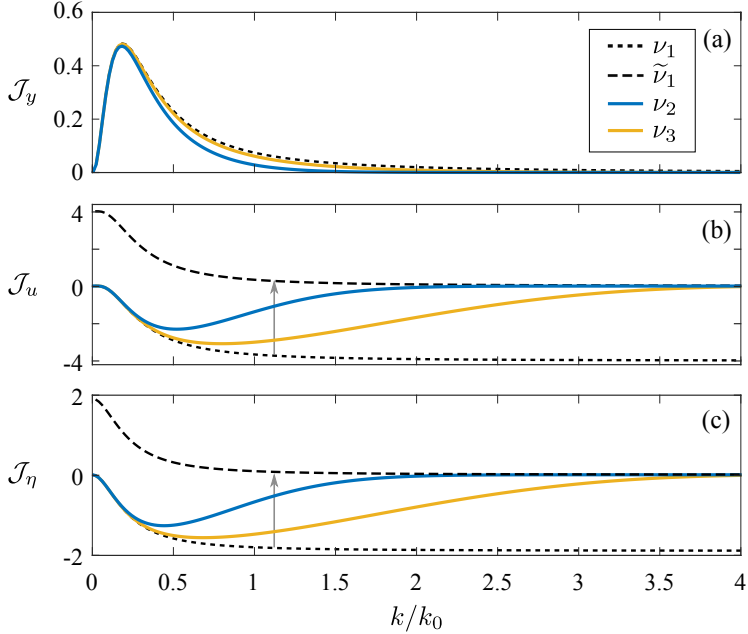}
\caption{Characteristic momentum dependencies of integrands entering equations for the density (a), chemical potential (b), and thermodynamic potential density (c), see also Eqs.~(\ref{eq:mu}), \eqref{eq:mu-renorm}, and \eqref{B.eq1}--\eqref{B.eq3}, at $\gamma=7.4\times10^{-4}$, $y=0.98$, $u=0.8$, and three different pseudopotentials with the fixed $k_0=0.105$. This case corresponds to $n=5\times 10^{15}$~cm$^{-3}$, $a_s^B=100a_0$, and $r_0=8$~nm. The arrows correspond to the renormalization procedure $\nu_1\to\widetilde{\nu}_1$ discussed in the text.}\label{fig2}
\end{figure}

As it is easy to see, the summand in Eq.~(\ref{eq:mu}) has the following asymptotic behavior at $p\to\infty$:
$$
	{\alpha_{\bf p}\over\omega_{\bf p}}-{gn_{0}\over 2\omega_{\bf p}}-1
    \approx -gn_{0}{m\over p^{2}}\,.
$$
Therefore, the corresponding integral diverges at the upper limit due to the factor $p^{2}dp$, see also Fig.~\ref{fig2}(b).
This divergence is associated with the fact that we have replaced the Fourier transform of the interaction potential $\nu({\bf p})$ by the constant value $g$. This difficulty also appears when calculating the ground state energy of a weakly interacting Bose gas [see, e.g., Refs. \cite{Pethick,Stringari}), and Fig.~\ref{fig2}(c)]. It can be overcome by the following renormalization of the coupling constant $g$ in the first term of Eq.~(\ref{eq:mu}):
$$
\nu_1({\bf p})\to \widetilde{\nu}_1({\bf p})=g+{g^{2}\over 2\mathcal{V}}\sum_{{\bf p}\neq 0}{2m\over p^{2}}.
$$
Thus, we arrive at
\begin{equation}\label{eq:mu-renorm}
	\widetilde{\mu}_1=gn_{0}+{g\over 2\pi^{2}\hbar^{3}}\int_{0}^{\infty}dpp^{2}
    \left[{\alpha_{\bf p}\over\omega_{\bf p}}
    -{gn_{0}\over 2\omega_{\bf p}}-1+gn_{0}{m\over p^{2}}\right].
\end{equation}
Now, the integral in Eq.~(\ref{eq:mu-renorm}) converges and must be computed in the range of physical parameters, where the quasiparticle spectrum is real. The attractive potential ($g<0$) does not satisfy this requirement at any momentum, while the repulsive interaction ($g>0$), according to Eq.~(\ref{eq:spectr}), yields the condition $\mu\leq gn_{0}$ that ensures the spectrum to be always real. Therefore, the quasiparticle description is only valid in the case of repulsive contact interaction potentials, moreover, the condition $\mu\leq gn_{0}$ must be satisfied. However, even in this case Eq.~(\ref{eq:mu-renorm}) {\it has no solutions}, since the corresponding integral is always positive and never turns to zero,  see Fig.~\ref{fig2}(b) for $\widetilde{\nu}_1$. Nevertheless, it should be noted that Eqs. (\ref{eq:2.24}), (\ref{eq:spectr}), and (\ref{eq:mu-renorm}) formally reproduce the well-known results for corrections of relevant quantities in terms of the gas parameter (see \ref{Ap}). 

We conclude that the point-like interaction potential leads to divergences in the quadratic approximation and even after the renormalization procedure aimed to remove them, the corresponding equation for a minimum of the thermodynamic potential is not satisfied. Therefore, it would be interesting to look at the problem by considering ``more realistic'' nonlocal potentials with nontrivial momentum dependencies of their Fourier transforms. It is worth stressing that the original Bogoliubov theory \cite{Bogoliubov} does not employ the contact interaction potential. Note also that the effects associated with a finite range of interaction potentials also occur in systems of atoms having dipole moments when the interaction is specified by the long-range dipole-dipole forces.

Therefore, as the second case, we consider the nonlocal interaction characterized by the Gaussian (normal) distribution, $V_2(r) = {g}{(2\pi r_0^2)^{-3/2}}\exp{(-
{r^2}/{2r_0^2})},$
where $r_0$ is the corresponding dispersion parameter that characterizes the spatial extent of the model potential, see also Fig.~\ref{fig1}. According to Eq.~\eqref{eq:nup}, the Fourier transform has the form
\begin{equation}\label{eq:nup2}
 \nu_2({\bf p}) = g\,e^{-(pr_0/\hbar)^2/2}.
\end{equation}

As the third case, we address the potential of semitransparent spheres with $V_3(r)=U_b$ at $r\leq r_0$ and
$V_3(r)=0$ at $r> r_0$ (see Fig.~\ref{fig1}), where the amplitude $U_b$ is convenient to express through the coupling constant $g$, $U_b=3g/(4\pi r_0^3)$. This leads to the Fourier transform of the form
\begin{equation}\label{eq:nup3}
 \nu_3({\bf p})  = 3g\frac{j_1(pr_0/\hbar)}{pr_0/\hbar},
\end{equation}
where $j_1(x)$ is the spherical Bessel function, $j_1(x)=\sin(x)/x^2 - \cos(x)/x$.
Note that the chosen model potentials $\nu_{2,3}$ in the limit $r_0\rightarrow0$, i.e., point-like objects, coincide with the contact interaction $\nu_{1}$, see also Fig.~\ref{fig1}. In this way, by allowing non-trivial momentum dependence of potentials of interacting particles, in addition to the spatial parameter $a_s$, it is required to introduce one more parameter $r_0$. As we discussed above, the characteristic spatial extent $r_0$ can not be neglected (i.e., simply set to zero) in the framework of the quadratic approximation.

To proceed with the numerical analysis, it is convenient to transform the obtained set of equations to dimensionless form. Replacing the summation by integration and using spherical coordinates, Eqs.~(\ref{eq:2.24})-(\ref{eq:2.26}) [see also Eqs.~(\ref{eq:2.11}) and (\ref{eq:2.15})] can be reduced to the following form, respectively,
\begin{eqnarray}\label{B.eq1}
 &&
 1=y+\frac{2\pi}{\gamma}\int_0^\infty k^2 d{k}
 \left[
  \frac{a(k)}{w(k)}-1
 \right],
 \\
 &&
 u=1+\frac{2\pi}{y\gamma}\int_0^\infty k^2 d{k} \left\{
  \frac{a(k)[1+s(k)] - bs^2(k)}{w(k)}-[1+s(k)]\right\},\label{B.eq2}
 \\
 &&
  \eta={y^{2}\over 2}-uy^{2}+{\pi^{2}\over\gamma^{2}}\int_{0}^{\infty}k^{2} dk\left[w(k)-a(k)\right], \label{B.eq3}
%
\end{eqnarray}
where $a(k)=k^2 + b[1-u+s(k)]$ and
\begin{equation}\label{wk}
 w(k)=\sqrt{\{k^2 + b[1-u+s(k)]\}^2 - b^2s^2(k)}
\end{equation}
is the dimensionless energy dispersion of the single-particle excitations, $\omega_{\bf p}=[\pi g/2(a_s^{B})^3]w(k)$. The scattering length in the first Born approximation,
\begin{equation}\label{as.v1}
	a_{s}^{B}={m\over 4\pi\hbar^{2}}\int d{\bf r}V({\bf r}),
\end{equation}
is directly related to the amplitude $g$ of both local and nonlocal potential $\nu({\bf p})$,
$g=4\pi\hbar^{2}a_{s}^{B}/m$,
and it is used as a scaling parameter in the numerical analysis. The relation between $a_s^B$ and the actual scattering length $a_s=a_s(g,r_0)$  is determined and discussed in the end of Sec.~\ref{sec4} for each model potential.

Now, the variables of interest are $y$, $u$, and $\eta$, where $y=n_0/n$ is the condensate fraction,
$u=\mu/(gn_0)$ is the dimensionless chemical potential ($u=1$ in the ``standard'' Bogoliubov approach),
and $\eta$ is the dimensionless thermodynamic potential, $\eta={\cal W}^{(2)}/gn^{2}$.
Among other parameters, $\gamma=n(a_s^B)^3$ can be related to the gas parameter ($\gamma\ll1$) and $b=2y\gamma/\pi$. The dimensionless integration variable $k$ is related to the particle momentum as $p=(2\pi\hbar/a_s^B)k$.
Therefore, the function $s(k)$ determines the spatial extent of the model potential in momentum space scaled in units of the dimensionless momentum $k$. In particular, for the contact interaction $s_1(k)=1$, for the Gaussian potential $s_2(k)=\exp[-k^2/(2k_0^2)]$, and for the semi-transparent spheres $s_3(k)=3j_1(k/k_0)/(k/k_0)$, where $k_0=a_s^B/(2\pi r_0)$.

From the point of view of ultracold-gases applications, there are only three spatial parameters that can be tuned: i) the interparticle distance related to the gas density~$n$, ii) the scattering length~$a_s$ adjusted by means of the Feshbach resonances, iii) the parameter~$r_0$ that determines the form and the spatial extent of the potential both in momentum and real spaces.
The two former are system-specific and the latter is the model-specific parameter. The spatial characteristics of the used model potentials in both real and momentum spaces are summarized in Fig.~\ref{fig1}.

In case of the model interaction potentials $V_2(r)$ and $V_3(r)$, let us point out physical limitations for the parameter $r_0$ that characterizes the spatial extent of the potential. First, this parameter is limited by a condition that it must exceed the typical atomic size of the order of the Bohr radius~$a_0$, where the repulsive part of the corresponding exact (or, e.g., Lennard-Jones) potential becomes very large. Second, it must characterize the two-body interaction processes only, i.e., not more than two atoms on average are allowed to enter the active region of $V(r)$, thus the extent $r_0$ must be limited from above by the average interparticle distance, $r_0\ll n^{-1/3}$. Therefore, we arrive at the following double condition:
\begin{equation}\label{eq:ineq}
 a_0\ll r_0\ll n^{-1/3}
\end{equation}
that must be fulfilled while employing $s_{2,3}(k)$.

Finally, let us discuss the analytic properties of functions in the integrals under consideration. It is natural to assume that $b>0$ and $w(k)\geq0$ are real.
The integral in Eq.~\eqref{B.eq1} converges for all pseudopotentials $s_i(k)$ under study. The main contribution to this integral is determined by the gas parameter $\gamma$ and only slightly affected by the choice of $r_0$,
unless it is taken too large or the condition $\gamma\ll1$ is no longer fulfilled.
At the same time, the integrals in Eqs.~\eqref{B.eq2} and \eqref{B.eq3} diverge for $s_1(k)$ in a similar manner (linearly), while for spatially-dependent potentials $s_{2,3}(k)$ the divergences at $k\rightarrow\infty$ are removed. The corresponding behavior of all integrands under study is summarized in Fig.~\ref{fig2}.

\section{Physical observables}\label{sec4}
\begin{figure}
\includegraphics[width=0.7\linewidth]{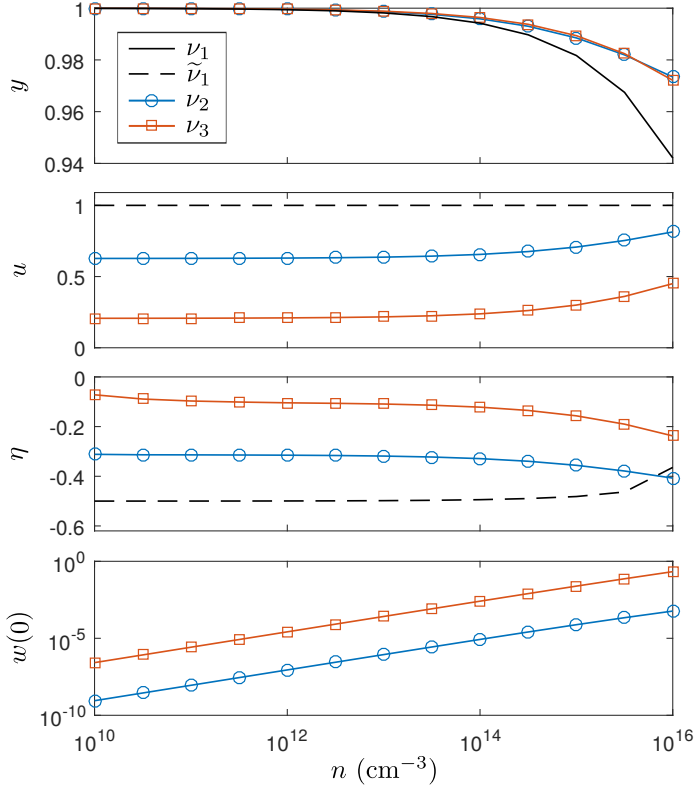}
\caption{Dependencies of the condensate fraction $y=n_0/n$, the chemical potential $u=\mu/gn_0$, the thermodynamic potential~$\eta={\cal W}^{(2)}/gn^{2}$, and the quasiparticle excitation gap $w(0)$ on the gas density~$n$ for three different pseudopotentials~$\nu_i$ at $a_s^B=100a_0$ and $r_0=8$~nm.}\label{fig3}
\end{figure}

To get an important physical insight, we start our analysis with the dependencies of the main observables such as the condensate density $n_0$, the chemical potential $\mu$, and the single-particle excitation gap $\omega_0$ on the gas density.
To proceed with the numerical approach at zero temperature, we choose the values for the scattering length~$a_s$ and the density~$n$ that are typical for experiments with ultracold dilute gases of alkali metal atoms. We also require the model parameter $r_0$ to fulfill the inequalities~\eqref{eq:ineq}. The corresponding results for the fixed $g$ that is parameterized by $a_s^B$, fixed $r_0$, and tunable gas density~$n$ are shown in Fig.~\ref{fig3}.

According to our analysis, the solutions suggest that the chemical potential itself differs from the value $\mu_0=gn_0$, thus leading to different results in physical observables. While the condensate density $n_0$ changes only quantitatively (it is systematically larger than in the ``standard'' Bogoliubov approximation with $\mu=\mu_0$), the single-particle excitation spectrum acquires a nonzero energy gap, $w(0)>0$. As one can see in Fig.~\ref{fig3}, the density of the thermodynamic potential $\eta$ is always negative (consequently, the pressure is positive) that indicates the stability of the system.

Let us note that the obtained results are sensitive to the choice of the spatial extent $r_0$ of the pseudopotentials under study. This becomes clear in the limit  $r_0\to0$ [although this limit is not ``inoffensive'' and violates the inequality~\eqref{eq:ineq} as discussed above], where both potentials $V_{2}(r)$ and $V_{3}(r)$ collapse to delta-functions.
As one can see from Fig.~\ref{fig2}, in the absence of the renormalization of integrals, both the chemical potential and the thermodynamic potential density diverge. In particular, according to Eq.~\eqref{eq:mu}, $\mu_1\to-\infty$ that results in divergence of the quasiparticle spectrum, i.e., the quasiparticle description  becomes invalid within the quadratic approximation.

\begin{figure}
	\includegraphics[width=0.75\linewidth]{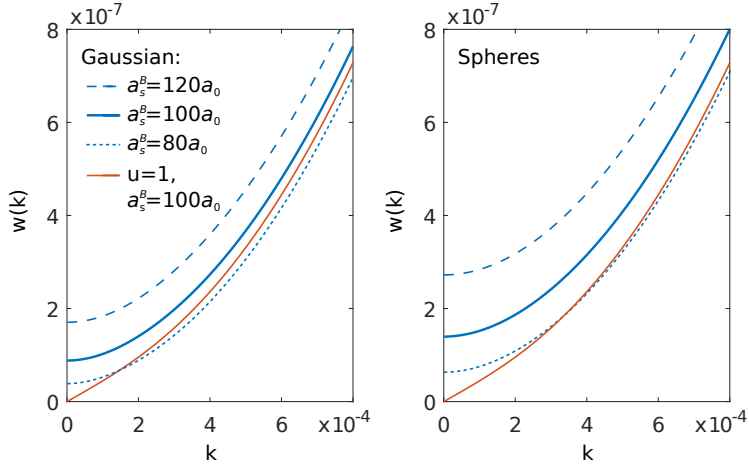}
	\caption{Quasiparticle energy dispersions for two model nonlocal potentials $V_2$ (left) and $V_3$ (right) at $r_0=8$~nm, $n=10^{12}$~cm$^{-3}$, and different amplitudes~$g$ that are parameterized by the corresponding $a_s^B$. 
    }\label{fig4}
\end{figure}

Next, we compare the quasiparticle energy dispersions obtained in the framework of the developed approach with the standard Bogoliubov approximation ($u=1$) for two model potentials $V_{2}(r)$ and $V_{3}(r)$. As it is shown in Fig.~\ref{fig4},
in addition to the nonzero energy gap of the order of $10^{-14}$~eV (that is approximately one order of magnitude smaller than the accessed values in the experimental measurements \cite{Steinhauer2002}), a non-linear character of the energy dispersion is observed at small momenta. 
Figure~\ref{fig4} also shows that with decrease of $g$ the gap $w(0)$ decreases. Numerical analysis indicates that in the limit $g\to0$ (with the fixed $r_0$) the contribution from the quadratic terms of the truncated Hamiltonian, i.e., the integral in Eq.~(\ref{B.eq2}), becomes small in comparison with the $c$-number term $u=1$. Therefore, in this limit the Bogoliubov gapless spectrum can be {approximately} recovered.
The spectrum is {\it exactly} gapless only for specific profiles of potentials $s({k})$ that set the integral in  Eq.~(\ref{B.eq2}) to zero.

As mentioned, the issue of the gapful spectrum of single-particle excitations has a long history and it has been debated in recent studies \cite{Suto2008,Bobrov2010,Ett,Poluektov2014}.  
The fact that there is a nonzero energy gap, in particular, seems to contradict the generally-accepted picture on spontaneous symmetry breaking of the continuous U(1) symmetry that must be accompanied by an appearance of the linear gapless Nambu-Goldstone collective (phonon) mode in the system with a BEC.
We discuss this issue in the last section.

Finally, we would like to address the issue of the scattering length for nonlocal potentials under consideration. In our study the amplitudes of nonlocal potentials $V_{2}(r)$ and $V_{3}(r)$ are parametrized by the coupling constant $g$ to recover the contact interaction $V_{1}({\bf r})=g\delta({\bf r})$ in the limit of zero range potentials ($r_{0}\to 0$). 
However, it is clear that for nonlocal potentials, the scattering length should be determined by both the interaction range and the amplitude of the potential. For a given nonlocal potential, the scattering theory of slow particles predicts, in principle, such a dependence. Although this is a rather complicated problem, it can be analytically solved for a number of model potentials. 

In particular, for the semitransparent spheres potential, $V(r)=V_{0}$ if $r\leqslant r_{0}$ and $V(r)=0$ if $r>r_{0}$, the phase shift determining the scattering length is given by \cite{Davydov},
\begin{equation*}
\delta_{0}=\arctan\left[k{\tanh(K_{0}r_{0})\over K_{0}}\right]-kr_{0}, \quad  \quad kr_{0}\ll 1,
\end{equation*}
where $k$ is the wave vector, $K_{0}^{2}=(2m_{*}V_{0}/\hbar^{2})$, and $m_{*}$ is a reduced mass. Then, the scattering length 
\begin{equation}\label{length_sph}
a_{s}=-{\tan\delta_{0}\over k}\approx r_{0}-{1\over K_{0}}\tanh(K_{0}r_{0}).
\end{equation}
For the potential~$V_3(r)$ used in the previous section, $K_{03}^{2}=3mg/4\pi\hbar^2r_{0}^{3}={3a^{B}_{s}/r_{0}^{3}}$. Therefore, Eq.~(\ref{length_sph}) sets the scattering length as a function of two parameters, $a_s=a_s(g,r_0)$, and it can be estimated in a straightforward way,
\begin{equation}\label{as.v3}
	a_s=
    r_0\left[1-\sqrt{r_0/3a_s^{B}}\tanh\left(\sqrt{3a_s^{B}/r_0}\right)\right].
\end{equation}
In particular, for the values used in the above numerical analysis,  $r_0=8$~nm~$\approx151a_0$ and $a_s^{B}=100a_0$, one obtains $a_s\approx56a_0$.

For the Gaussian potential of the form
$V(r)={V_{0}\over 2L^{2}}e^{-{r^{2}\over L^{2}}}$,
the scattering length as a function of $V_{0}$ and $L$ was given in Ref.~\cite{Jeszenszki2018},
\begin{equation}\label{length_gauss}
{a_{s}\over L}={\sqrt{\pi}\over 2}{V_{0}\over{V_{0}+{2\hbar^{2}\over m_{*}}}}.
\end{equation}
Comparing this potential
with $V_{2}(r)$, we have $L\equiv \sqrt{2}r_0$, $V_0\equiv g\sqrt{2}/(\pi^{3/2} r_0)$ and, consequently, Eq. (\ref{length_gauss}) becomes 
\begin{equation}\label{as.v2}
	a_s=r_0\frac{\sqrt{\pi/2}}{1+ \sqrt{\pi/2}\cdot r_0/a_s^{B}}.
\end{equation}
Again, taking the values $r_0=8$~nm~$\approx151a_0$ and $a_s^{B}=100a_0$ used in the numerical analysis, we obtain $a_{s}\approx 65a_{0}$.

Both estimated values of $a_s$ correspond to the moderate regime of interactions in dilute gases of alkali-metal atoms. Hence, while estimating the actual values of the scattering length~$a_s$ relevant for experiments, it is necessary to use Eqs.~\eqref{as.v3} and \eqref{as.v2} for the corresponding nonlocal potentials.

\section{Summary and discussion} \label{sec5}

We derived the general coupled equations of quadratic approximation for a weakly interacting Bose gas by using the Bogoliubov method based on extracting the condensate modes and subsequent diagonalization of the truncated Hamiltonian that is quadratic in creation and annihilation operators of non-condensate particles. They include the equation that provides a relation between the total particle number and chemical potential, as well as  equation for the condensate density as a variational parameter. The latter ensures the minimum of the grand thermodynamic potential. 

In contrast to the standard approaches, where the contribution of the quadratic terms to the  equation that guarantees a minimum of the thermodynymic potential is usually neglected \cite{Dalfovo,Pethick,Stringari} (thus, the equations become decoupled), in this study we, for the first time, solved the coupled equations \emph{self-consistently} for a range of model potentials that have well-defined analytic expressions in both real and momentum spaces.
For the chosen nonlocal potentials, we found that the obtained solutions do not depend qualitatively on the shape of the spatial distributions of the potential. Therefore, we also believe that they will be qualitatively similar for other nonlocal potentials. The main results are summarized as follows:

\begin{enumerate}
	\item The integrals in general equations of quadratic approximation diverge at large momenta for the contact interaction potential. Even after the standard renormalization procedure aimed to remove them, the corresponding equations have no solutions, although they formally reproduce the well-known results such as corrections to the chemical potential and condensate density in terms of the gas parameter (see \ref{Ap}).
	\item In the case of nonlocal interactions that are repulsive and contain a contact interaction as a limiting case,  all the integrals under study converge and the coupled equations themselves have nontrivial solutions for realistic parameters of atomic gases.  The contribution of the terms originating from the quadratic part of the truncated Hamiltonian can be of the same order as from those coming from its $c$-number part. This yields a nonzero gap in the spectrum of the single-particle excitations, in contrast to the situation with the renormalized contact interaction potential.
	\item The structure of the coupled equations and analyzed solutions suggest applicability of the quadratic approximation to a wide range of nonlocal potentials including those that can have both attractive and repulsive regions in real space (see \ref{Ap2} for details). 
    \item The numerical analysis involving nonlocal potentials leads to a conclusion that the results are sensitive to their choice only quantitatively, but not qualitatively. Therefore, one might expect that the qualitative side of this study is valid for more realistic physical potentials. The main limitation in using these (e.g., Lennard-Jones) potentials in the framework of the developed approach concerns only the existence of the corresponding Fourier transforms, i.e., accurate representation of the interatomic potential in the momentum space.
\end{enumerate}

Experimental studies both involving ultracold gases of alkali-metal atoms with BEC \cite{Steinhauer2002} and superfluid helium \cite{Beauvois2018} point to the relative agreement of the observations with the predictions based on the original Bogoliubov approach. However, the region of small momenta (where the quadratic approximation for nonlocal potentials points to the qualitatively different behavior with the gap at $p=0$, see Fig.~\ref{fig4}) was not directly accessed there so far, thus an unambiguous experimental proof of either presence or absence of the gap in the single-particle excitation spectrum is still missing.

In connection with the Goldstone theorem, the inevitable question arises concerning the existing gap in the excitation spectrum of a weakly interacting Bose gas obtained by diagonalizing the truncated Hamiltonian. It is generally accepted that the Bogoliubov spectrum of the single-particle excitations describes the collective Goldstone mode associated with the broken U(1) symmetry, i.e., the energy of these excitations vanishes at zero momentum. This originates from the fact that the Bogoliubov gapless spectrum, in its original form, can be only obtained for the chemical potential $\mu=\nu(0)n_{0}$ satisfying the Hugenholtz-Pines relation \cite{Hugenholtz}. Moreover, the spectrum is determined by the poles of the single-particle Green's function shared with those of two-particle Green's function as was first found by Gavoret and Nozi\`{e}res \cite{Gavoret}. The poles of the former specify the single-particle excitations, while those of the latter specify the corresponding collective excitations (phonons).   

However, as it is shown in the present study, for non-local potentials the chemical potential is no longer determined as $\mu=\nu(0)n_{0}$, thus it does not meet the Hugenholtz-Pines relation \cite{Hugenholtz}. This is due to the fact that the contribution from the quadratic terms entering the truncated Hamiltonian to the chemical potential is of the same order of magnitude as from the $c$-number terms and, thus, must be taken into account. In addition, a thorough analysis of field-perturbative expansions for the single- and two-particle Green's functions performed recently by Kita \cite{Kita2009,Kita2010} showed that their poles are not shared, contrary to the conclusion made by Gavoret and Nozi\`{e}res \cite{Gavoret}. Moreover, the subsequent study confirmed the different character of single-particle and collective excitations \cite{Kita2011,Kita2016}. In particular, the width of the collective-mode spectrum manifestly vanishes in the long-wavelength limit, whereas that of the quasiparticle spectrum (single-particle excitations) apparently remains finite \cite{Kita2016} and the Goldstone mode can appear as a pole of the two-particle Green's function \cite{Kita2011}. The similar findings  concerning the different nature of excitations in an interacting Bose system with BEC were recently reported in Refs.~\cite{Bobrov2015,Bobrov2016}. Therefore, the spectrum of interacting Bose system may consist of two branches \cite{Ett,Bobrov2010,Bobrov2016,Poluektov2014}, similarly to the BCS theory of neutral superfluids. The first branch represents the single-particle excitations that can be gapful. The second one characterizes the gapless collective mode (density oscillations). The separation of the single-particle and collective excitations is probably less manifested in a Bose system than in a Fermi system as a consequence of the hybridization of branches due to the presence of a condensate.    
Therefore, the nature of the excitation spectrum of interacting Bose system with BEC can be more complex than it is commonly believed and additional advanced theoretical and experimental investigations in this direction are required.

\begin{appendix}
\section{Quadratic approximation for the renormalized contact interaction}\label{Ap}

The equations obtained in the main part formally reproduce the known results (see, e.g, Refs.~\cite{Dalfovo,Pethick,Stringari}). Indeed, if one considers $\mu=gn_{0}$ as the leading term originating from the $c$-number part of the truncated Hamiltonian, the single-particle excitation spectrum becomes gapless and acquires the Bogoliubov form,
$$
\omega_{\bf p}=\left[\left({p^{2}\over 2m}\right)^{2}+{p^{2}\over m}gn_{0}\right]^{1/2}.
$$
The substitution of $\mu=gn_{0}$ and $\omega_{\bf p}$ into the left-hand side of Eq. (\ref{eq:mu-renorm}) with subsequent integration yields (here $\tilde{\mu}_{1}\equiv\mu$)
\begin{equation}\label{eq:A1}
\mu=gn_{0}\left[1+{40\over 3\sqrt{\pi}}\sqrt{n_{0}a_{s}^{3}}\right].
\end{equation}
In a similar manner, the straightforward integration of Eq. (\ref{eq:2.24}) for $\nu({\bf p})=g$ gives the total particle density,
\begin{equation}\label{eq:A2}
n=n_{0}\left[1+{8\over 3\sqrt{\pi}}\sqrt{n_{0}a_{s}^{3}}\right].
\end{equation}
Next, using Eqs. (\ref{eq:A1}) and (\ref{eq:A2}) one obtains
\begin{equation}\label{A3}
\mu=gn\left[1+{32\over 3\sqrt{\pi}}\sqrt{n_{0}a_{s}^{3}}\right].
\end{equation}
The latter equation provides a relation between the chemical potential and total particle density. The second term in the right-hand side of Eq.~(\ref{eq:A2}) describes the so-called quantum depletion of the condensate. It is expressed in terms of the gas parameter $\gamma=na^{3}$ that is required to be small, $\gamma\ll 1$ ($n_{0}\approx n$). The terms containing the gas parameter originate from the quadratic part of the truncated Hamiltonian and are usually treated as the corrections to the corresponding quantities.

\section{Analysis for the model potential with attractive regions in real space}\label{Ap2}

In connection with the realistic interatomic potentials (e.g., of the Lennard-Jones type) in weakly-interacting gases with BEC, the natural question arises whether the results of the developed approach change qualitatively when the nonlocal model potential consists of both the repulsive and attractive regions in real space.
From the brief analysis of the coupled equations~\eqref{eq:2.24} and \eqref{eq:2.25} we arrive at the general conclusion that this should not be the case (the solutions depend rather on the asymptotic behavior of the model potentials in momentum space at $p\to\infty$).

To be more specific and provide some quantitative estimates for the sign-changing interaction, we perform an additional analysis for the following potential:
\begin{equation}\label{eq:Vr4}
 V_4({\bf r})  = \frac{g}{2\pi^2 r_0^3}\frac{j_1(r/r_0)}{r/r_0},
\end{equation}
which, by inverting the case of semi-transparent spheres, see Eq.~\eqref{eq:nup3} and Fig.~\ref{fig1}, has the Fourier transform
\begin{equation}\label{eq:Vp4}
 \nu_4({\bf p})  = g\theta(\hbar/r_0 - p),
\end{equation}
where $\theta(x)$ is the Heaviside step function. The potential $V_4({\bf r})$ has both repulsive and attractive regions (see also Fig.~\ref{fig1}; however, $V_4({\bf r})$ can not be normalized, in contrast to $V_2({\bf r})$ and $V_3({\bf r})$, that was used to directly relate to the limit of delta-function) and its Fourier transform $\nu_4({\bf p})$ has proper asymptotic behavior at $p\to\infty$ to ensure convergence of the integrals that enter Eqs.~\eqref{B.eq1}-\eqref{B.eq3}.

\begin{figure}
\includegraphics[width=0.7\linewidth]{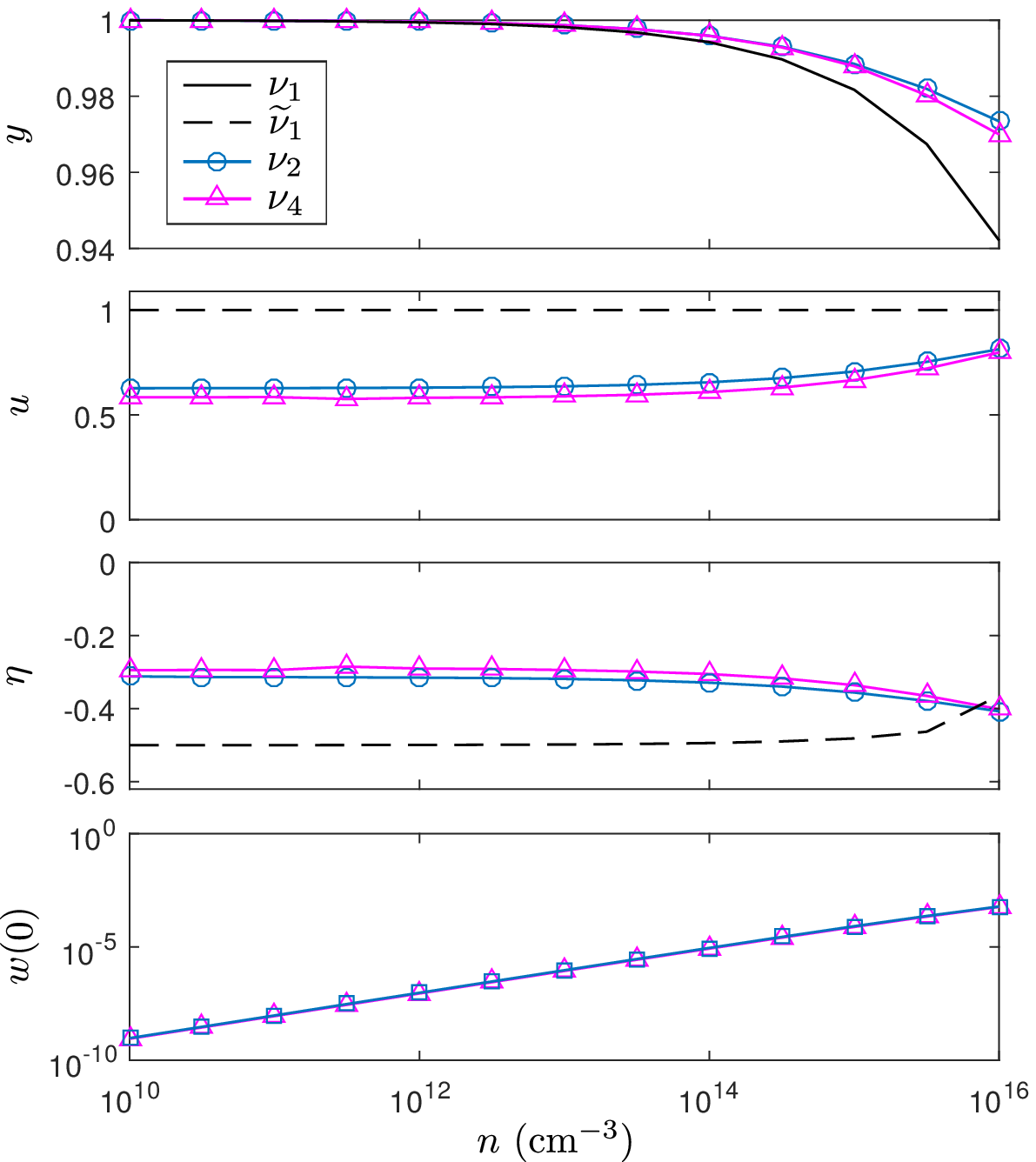}
\caption{
	Dependencies of the condensate fraction $y=n_0/n$, the chemical potential $u=\mu/gn_0$, the thermodynamic potential~$\eta={\cal W}^{(2)}/gn^{2}$, and the quasiparticle excitation gap $w(0)$ on the gas density~$n$ for three different pseudopotentials~$\nu_i$ at $a_s^B=100a_0$ and $r_0=8$~nm.
}\label{fig5}
\end{figure}

In Fig.~\ref{fig5} we show the results of the corresponding numerical analysis for this model potential. They confirm the above statement that the shape of the nonlocal interaction potential has only quantitative effect on the physical observables under study. In particular, for the given parameters $a_s$ and $r_0$ the results for the potential $V_4({\bf r})$ resemble the case of the potential with the Gaussian profile $V_2({\bf r})$ up to minor corrections.

Yet another model potential $V_5(r)=-({g}/{8\pi r_0^3})e^{-r/r_0}$ is sometimes used in the ultracold gases literature \cite{Caballero2013}, in particular, to reproduce an attractive tail of interaction. This potential has well defined analytic expressions for the corresponding scattering length, $a_s=a_s(g,r_0)$ \cite{Rarita37}, and its Fourier transform, $\nu_5(p)={-g}/[1+(pr_0/\hbar)^2]^2$. However, it can not be solely used in the framework of the developed approach at $g>0$. 
According to Eq.~(\ref{eq:2.15}), the quasiparticle energy becomes complex quantity at some values of the momentum~$p$. This means that the quasiparticle description becomes invalid. Nevertheless, in a linear combination with (another) positive-defined potential (reproducing the repulsive part of the corresponding Lennard-Jones potential; e.g., $V_{2,3}(r)$) it can be used in the regimes when the resulting quasiparticle spectrum remains real. This is a perspective direction, but due to the enlarged number of tunable parameters, goes beyond the scope of the current study.

\end{appendix}

\section*{Acknowledgements}
A.S. acknowledges funding from the European Research Council (ERC) under the European Union's Horizon 2020 research and innovation programme (grant agreement No. 646807-EXMAG).

\section*{References}
\bibliography{quadr_approx} 

\end{document}